\documentstyle[12pt]{article}
\begin{document}

\def\br{\begin{eqnarray}}
\def\er{\end{eqnarray}}
\def\be{\begin{equation}}
\def\ee{\end{equation}}
\def\a{\alpha}
\def\b{\beta}
\def\c{\chi   }
\def\d{\delta}
\def\D{\Delta}
\def\eps{\epsilon}
\def\g{\gamma}
\def\G{\Gamma}
\def\grad{\nabla}
\def\h{ {1\over 2}  }
\def\hp{ {+{1\over 2}}  }
\def\hm{ {-{1\over 2}}  }
\def\i{\int}
\def\k{\kappa}
\def\l{\lambda}
\def\L{\Lambda}
\def\m{\mu}
\def\n{\nu}
\def\o{\over}
\def\O{\Omega}
\def\p{\phi}
\def\pa{\partial}
\def\pp{ p^+}
\def\pr{\prime}
\def\r{\right}
\def\ra{\rightarrow}
\def\rh{\rho}
\def\s{\sigma}
\def\sp{\sigma^+}
\def\t{\tau}
\def\th{\hat t}
\def\sh{\hat s}
\def\x{\tilde x}
\def\df{\frac{\chi^{2}}{d.f.}}
\def\v{\vert}
\def\qb{\bar q}
\def\b{\tilde b}
\def\inte{\int dx }
\def\integ{\iint dt dx }
\def\lie{{\cal G}}
\def\dlie{{\cal G}^{\ast}}
\def\hlie{{\widehat {\cal G}}}
\def\({\left(}
\def\){\right)}
\def\Tr{\mathop{\rm Tr}}
\def\myone{1 \hspace{-.7mm}\rule{.26mm}{2.7mm}}

\title
\large {\bf A QUANTITATIVE INVESTIGATION OF THE POMERON }

\normalsize
\vskip .4in

\begin{center}
 E. Gotsman $^{(a),} $\footnotemark
\footnotetext{Work supported in part by Israel Academy of Sciences
and Humanities}E.M. Levin $^{(b)}$ and U. Maor$^{(a),}$
 \footnotemark \footnotetext{ Work supported in part by the MINERVA
 Foundation}
\par \vskip .1in \noindent
$ ^{(a)}$ School of Physics and Astronomy \\
Raymond and Beverly Sackler Faculty of Exact Sciences\\
Tel Aviv University, Tel Aviv\\
\par \vskip .1in
 \par \vskip .1in \noindent
$^{(b)}$ DESY, Hamburg, Germany \\ and \\
St. Petersbourg Nuclear Physics Institute,
Gatchina, Russia \\
\par \vskip .1in

\end{center}

\par \vskip .3in

\begin{abstract}
\par \vskip .3in \noindent
A comparative investigation of various Pomeron models is carried out
through the study of their predicted values of $ \sigma_{tot}$,
B, and $\frac{\sigma_{el}}{\sigma_{tot}}$ in high energy pp and
p$\bar{p}$ scattering. Our results strongly support a picture of
 the Pomeron in which we have both moderate blackening and expansion
of the p($\bar{p}$) - p amplitude in impact parameter space as a
function of energy in the ISR-SSC domain. In particular, we
obtain an excellent reproduction of the data with a hybrid
                       eikonal model
 which combines the hard Lipatov-like QCD Pomeron with the old
fashioned soft Pomeron and Regge terms.
Our analysis shows that the additive quark model, at least in the
naive form, is not compatible with the data.
\end{abstract}
%
\newpage

 At present there are two incongrous descriptions of the Pomeron.
One is that based on the conventional  phenomenology of soft hadronic
scattering  and the relevant calculations are carried out
  \cite {KTR}   within the framework of
Reggeon field theory .
The soft Pomeron is constructed from multiperipheral hadronic (Regge)
exchanges and has $\alpha_{P}$(0) =1. This is not compatible with the
high energy experimental observation of rising hadronic cross sections.
The old fashioned soft Pomeron was, thus replaced \cite{KTR,LA}
by a soft supercritical  Pomeron with an intercept
$ \alpha_{P}$(0) -1 $ > $ 0   .
 This model provides a reasonable description of pp and p${\bar p} $
scattering in and above the ISR energy range   .
 It's principle assumption, that the supercritical Pomeron is a simple
isolated pole in the complex angular plane, lacks field theoretic
justification, and contradicts our experience with the
multiperipheral model which fails to produce a pole
with $ \alpha_{P} \,>$ 1 .

\par
        The alternate description of the Pomeron is within
the framework of QCD.
Lipatov \cite{LIP} has shown that perturbative QCD, in which a hard
Pomeron is built out of multiperipheral high transverse momentum
gluon exchanges, predicts a different
Pomeron i.e. a series of poles in the complex j plane above unity at
1 $< j \, <\,  1 + \epsilon$ . The resulting Lipatov Pomeron has
a more complicated form $  \frac{s^{\alpha}}{(lns)^{\beta}}  $
with $ \alpha > 1  $ .
   A consequence of a pole with  intercept
 $ \alpha(0)$  is that $\sigma_{tot} \propto s^{\alpha(0)-1} $.
Hence, both the soft supercritical and the hard QCD Pomeron
 predict a powerlike rise of the total hadronic cross section.
This may continue until the unitarity bound gives rise to
 screening effects which saturates
   the growth of the total cross section \cite {LR}
making $ \sigma_{tot} \leq ln^{2} s $, which is compatible with the
the Froissart limit \cite {FR}.

\par
In the following we shall examine a wide class of models of the Pomeron.
Our goal is to compare the predictions of these models with the high
energy experimental data presently available \cite{DATA} over
the  energy range 23 $ \leq \sqrt{s} \leq $  1800 GeV.
  The purpose of such a comparative study is to attempt to
discriminate between models with different high energy
predictions based on data which are apparently \cite {BHM}
  below asymptotia.

\par
   A quantitative investigation of the various Pomeron models can be
 readily carried out in the impact parameter space.
Our amplitudes are normalized such that
 $ \frac{d \sigma}{dt} = \pi \vert f(s,t) \vert ^{2} $
; $ \sigma_{tot} = 4 \pi$ Im f(s,0)
 where
 \begin{equation}
 f(s,t) = \frac{1}{2 \pi} \int d{\bf b} e^{i{\bf q.b}} a(b,s)
\end{equation}
 and
 \begin{equation}
  a(s,b) = \frac{1}{2 \pi} \int d{\bf q} e^{-i{\bf q.b}}f(s,t)
\end{equation}
Hence we have   $ \sigma_{tot} = 2 \int d{\bf b}$ Im a(s,b)
and $ \sigma_{el} = \int d{\bf b} \vert a(s,b) \vert^{2} $ .
 The corresponding forward slope and curvature parameters are
 \begin{equation}
 B =\frac{ \int d{\bf b} b^{2} a(s,b)}{2 \int d{\bf b} a(s,b)} .
\end{equation}

 \begin{equation}
\nonumber
C= \frac{1}{32}
 \frac{ \int d{\bf b} b^{4} a(s,b)}{ \int d{\bf b} a(s,b)}  -
 \frac{1}{16} \mid \frac{ \int d{\bf b} b^{2} a(s,b)}
 { \int d{\bf b} a(s,b)} \mid^{2}
\end{equation}
where $$ \frac{d\sigma}{dt} = [\frac{d\sigma}{dt}]_{t=0} \;\;
  e^{( Bt + Ct^{2} +......)} $$

\par
  Unitarity requires  Im a(s,b)$ \leq 1 $.
In order to satisfy this  constraint  it is convenient
to express a(s,b) in terms of the complex eikonal function
$ \chi (s,b) $ with
\begin{equation}
 a(s,b) =i [1\,  - \, e^{i \chi(s,b)}]
\end{equation}
where  Im $ \chi  \geq $ 0.
This ensures that unitarity is restored on summing  all the
eikonal multi-particle exchange amplitudes.
  At high energy, forward elastic scattering is essentially diffractive
and therefore $ Re \chi $ is very small.
  We  assume that $ Re \chi \approx $ 0, consequently
 the amplitude a(s,b) is purely imaginary and determined
by the opaqueness $ \Omega (s,b) = Im \chi $ .
This assumption limits our study to the forward direction,
where the bulk of the experimental data is present.
\par We would like to note that considering only the
 pure imaginary $\Omega (s , b)$
violates the crossing symmetry of the scattering amplitude. We will
describe later how we restore this symmetry, assuming that
the real part of the amplitude is small enough.
One of our major objectives is to critically examine how much of the
observed change in the forward high energy pp and p${\bar p}$
scattering amplitudes can be associated with blackening
 as opposed to the expansion
of $ \Omega(s,b) $ and a(s,b) with s. We note that a soft Pomeron
trajectory is essentially flat i.e. $ \alpha^{\prime} $ is either zero
 or very small. Accordingly, the growth of the cross section  associated
with the soft Pomeron is attributed mostly to a blackening of the b-space
 scattering amplitude. This property is common to many other models
\cite{BSW,DP}. The b-space characteristics  of the Lipatov Pomeron
are less clear. In the LLA one expects \cite{LR,LERY} a moderate
expansion with energy. We will check whether this result is compatible
with data.
\par
Even though we choose an eikonal approach we do not attach
 any profound theoretical importance to this choice.
  It is merely a method to
restore  s-channel unitarity, and it enables us to assess the
    natural scale
of the shadowing corrections. These can manifest themselves either
on the hadronic level with a typical scale $R_h\approx 1 fm $ or on
the quark scale with a typical value of  $ R_Q \approx 0.1 fm$.
We expect the data analysis to provide us with some information on this.
 We would like to mention here that the  eikonal formula (5)
 allows us to
  incorparate in our
 fitting procedure new ideas, such as were discussed in refs.
                                            \cite{LERY}
 , \cite{RY} about the shadowing correction using different assumption
about $\Omega (s , b )$.
\par
To this end we compare several  Pomeron models with three
experimental quantities, which are particularly sensitive to the above,
i.e. $ \sigma_{tot}$ , B and $ \frac{ \sigma_{el}}{ \sigma_{tot}}$ .
We note that more than $\sigma_{el}$ itself, the ratio
 $ \frac{\sigma_{el}}{ \sigma_{tot}} $ is sensitive to the hadron
opacity. Over the ISR-Tevatron energy range
 this ratio was found \cite{DATA} to be in the interval
 0.15- 0.25, which is well below the saturation
 limit of 0.5 corresponding to a fully absorbing black disc \cite {BHM}.
 \par
The above experimental quantities relate directly to the structure
of the inelastic (multiparticle) high energy amplitude. Indeed,
$\sigma_{tot}$  is driven by the inelastic amplitude, B is the
mean radius of the  partonic distribution in the impact parameter
space and $\frac{\sigma_{el}}{\sigma_{tot}}$  characterizes  the
hadronic  opaqueness  which is very sensitive to the scale of the
shadowing corrections.
\par
    We examine the entire statistics of 74 data points \cite {DATA}
 of $ \sigma_{tot}$, B and $ \frac{\sigma_{el}}{ \sigma_{tot}} $
obtained from to pp and p${\bar p}$ scattering in the ISR-Tevatron
energy range. In our handling of the data, we have used the
published results as such, without any attempt to discriminate
or average between different ISR data sets at the same energy, or
re-analyse the raw data as recently advocated \cite {HM}.
In particular, our initial supposition, that a(s,b) is imaginary,
prevents us from  calculating  $ \rho = \frac{Re f}{Im f} $
without additional assumptions.

\par
 The data we have examined are characterized, in general, by relatively
 small errors. Consequently, one must carefully evaluate the
quality of the fit to this data. In our study we have found that
seemingly good reproductions of this data set, yield an unexceptable
 $ \frac{\chi^{2}}{d.f.}  > $ 10. An additional fitting problem
is associated with the fact that most of our data comes from
 the ISR, with the UA4 and Tevatron providing only 6 high energy
 points. As a result we had to check, that fits with an acceptable
$ \frac{\chi^{2}}{d.f.} $ also had a reasonable high energy behaviour.
We would also like to mention that the detailed high t structure of
the  elastic cross section is beyond the scope of our investigation.
This  corresponds to the very fine details of $a (b,s) $ or
$\Omega (b,s)$ at small b, a region which we do not
pretend to investigate.
\par
   We have attempted a number of generic forms for the Pomeron,
with different degrees of complexity. As we are discussing both
 pp and p${\bar p}$ scattering, we have also included in our
parametrization , an odd Regge term to account for the p${\bar p}$
and pp difference.
We therefore consider an amplitude of the form
\begin{eqnarray}
a(s,b) = A_{P}(s,b) \pm A_{R}(s,b)
\end{eqnarray}
Initially, we attempt an orthodox non-unitarized parametrization
of the Pomeron. This may possibly be adequate, since the measured data
is well below the unitarity limit, and there may be no  need to use
unitarized expressions in our analysis.
At this level we  have
for the soft Pomeron and Reggeon
\begin{eqnarray}
F_{i}(s,t) = c_{i} e^{R_{0i}^{2}t} s^{\alpha_{i}(t) -1}
sin[\frac{\pi}{2} \alpha_{i}(t)]
\end{eqnarray}
where i = P or R. With linear trajectories
 $ \alpha_{i}(t) = \alpha_{i}(0) + \alpha_{i}^{\prime} $ t,
 we can readily take the Fourier transform of Eq.(7) and obtain \\
$  A_{i}(s,b) \,\, = $
\begin{eqnarray}
             c_{i} \frac{s^{\alpha_{i}(0)-1}}{\mid\beta_{i} \mid^{2}}
exp[- \frac{b^{2}}{4 \mid\beta_{i} \mid^{2}}R_{i}^{2}] \cdot
(R_{i}^{2} sin[ \frac{\pi}{2} \alpha_{i}(0) + Z_{i}]
   - \frac{ \pi}{2} \alpha_{i}^{ \prime} cos[ \frac{ \pi}{2} \alpha_{i}
(0) + Z_{i}])
\end{eqnarray}
where
\begin{eqnarray}
R_{i}^{2} = R_{0i}^{2} + \alpha_{i}^{ \prime} lns
\end{eqnarray}
$$ \mid\beta_{i} \mid^{2} = R_{i}^{4} + \frac{ \pi^{2}}{4} $$
$$Z_{i} = \frac{ \pi \alpha_{i}^{\prime} b ^{2}}{8 \mid\beta_{i}\mid^{2}}
$$
Accordingly, $ A_{i}(s,b) $ has four adjustable parameters :
 $ c_{i}, R_{0i}^{2}, \alpha_{i}(0)$ and $ \alpha_{i}^{\prime} $.

\par
 The first model that we have tested has a supercritical soft Pomeron,
 and we denote it as $ a_{I}$ (s,b). Such a pomeron can be viewed
  as the consequence of the soft multihadron (multiparton)
 production at high energies.       The best fit parameters and the
 $ \frac{\chi^{2}}{d.f.}$ for this model are summarized in the upper
 section of Table I. In our original fit we found that
$ \alpha_{P}^{\prime} \approx$ 0.3, $ \alpha_{R}^{\prime} \approx $
 1. and $ \alpha_{R}(0) \approx $ 0.5. We have, therefore, fixed
these parameters at their traditional values to somewhat improve our
$ \df$ .
  The $ \frac{\chi^{2}}{d.f.} $ = 2.04 is misleading.
The model, as such, provides a good description of the ISR data at the
expense of a very poor reproduction of the UA4 and Tevatron data.
 This problem was traced to a low fitted value of
 $ \epsilon  =  \alpha_{P} $ - 1. We have also tested a model in
which $ \epsilon $ is fixed at 0.085 and denote it
as $ a_{II}$ (s,b,$\epsilon$ = 0.085).
The quality of the fit, presented in the upper section of Table I,
is considerably poorer, but yields a fair continuation from
the low to the high energy domains.

\par
   The third model we have tested is that based on the Lipatov-like
hard Pomeron, where we set
\begin{eqnarray}
A_{L1}(s,b)= \frac{a_{1}s^{a_{2}}}{(lns)^{a_{3}}}
\exp[- \frac{b^{2}}{R_{L1}^{2}(s)}]
\end{eqnarray}
with
\begin{eqnarray}
R_{L1}^{2}(s) = a_{4} + a_{5} (lns)^{a_{6}}
\end{eqnarray}
The $ a_{i} $ denote six adjustable parameters.
The parametrization considered is
\begin{eqnarray}
a_{III}(s,b) = A_{L1}(s,b) \pm A_{R}(s,b)
\end{eqnarray}
This parametrization assumes the soft contribution to be entirely
associated with the Regge sector whereas the hard contribution
is given by the Lipatov-like Pomeron driven by hard multihadron
 (multiparton) production.
\par
The form we have choosen for $ A_{L1} $ is very similar in it's
energy dependence to the form suggested by Bourrely, Soffer and Wu
\cite{BSW}. The two models are, however, fundamentally different.
 In the Bourrely, Soffer,Wu model, A(s,b) has a fixed b distribution
independent of s. Accordingly, the rise in $ \sigma_{tot} $
 is due only to the blackening of A(s,b). The Lipatov 1 amplitude that we
have examined, e.g. Eq. (10), has a b distribution which is energy
dependent, and hence $ R_{L1}^{2} $ is s dependent, see Eq. (11).

\par
Our best fit with the corresponding $ \frac{\chi^{2}}{d.f.} $ is
summarized in the upper section of Table I. As can be seen, the
hard Lipatov-like Pomeron provides an excellent fit, which is
considerably better than the soft Pomeron descriptions. However, it is
important to note that neither of these models are suitable
parametrizations at very high energies as the power dependence of
 s will eventually violate unitarity. This is particularly true
for the Lipatov  $ A_{L1}(s,b) $ where we have
 $ A_{L1}(s,b=0) \, >$ 1  for $ \sqrt{s}\,  >$  62 TeV.

\par
To overcome this problem we have also examined the above three
Pomeron models in an eikonal parametrization, where we set
\begin{eqnarray}
\Omega_{I}(s,b) = A_{P}(s,b) \pm A_{R}(s,b)   \\
\Omega_{II}(s,b,\epsilon = 0.085) =
 A_{P}(s,b,\epsilon = 0.085) \pm A_{R}(s,b)   \\
   \Omega_{III}(s,b) = A_{L1}(s,b) \pm A_{R}(s,b)
\end{eqnarray}
Our motivation for utilizing the eikonal expansion rests, not only on
the need to unitarize our amplitudes  and study the shadowing effects
                                    , but also on the knowledge
 \cite{CYY} that the eikonal description provides a natural
explanation for the changing logarithmic slope of
$ \frac{d \sigma}{dt} $ with t.
Some prudence must be exercised at this stage. Our investigation aims
at understanding
  the general characteristics of the Pomeron. As is well known,
neither of Eqs.(13 - 15) can reproduce  the dip structure at ISR and
its slow change to a shoulder as observed by UA4. However, these are
exeedingly small phenomena where
                                $\frac{d\sigma}{dt}$
                                                     at the second
peak is $ 10^{-5} - 10^{-6}$ smaller than
                                $\frac{d\sigma}{dt}$
in the forward direction. Accordingly, when viewed in $b$ - space it
corresponds to a very subtle substructure of $\Omega (b,s)$ at
small $b$, which is completely outside the scope of our investigation.
\par
  Our best fits and the corresponding $\frac{ \chi^{2}}{d.f.} $
   are presented in the middle section of Table I.  We  again
note, that the $ \Omega_{I} $ fit agrees well with the ISR
data ,
 but fails to reproduce the high energy data, due to the
the low value of $\epsilon$  in this fit.
  The quality of $\Omega_{III}$ fit is considerbly worse than the one
 obtained without eikonalization. We have therefore attempted a
  hybrid model
\begin{eqnarray}
\Omega_{IV}(s,b) = A_{L1}(s,b) + A_{P}(s,b) \pm A_{R}(s,b)
\end{eqnarray}
Namely, we associate the increase of $ \sigma_{tot} $ with the hard
QCD Lipatov-like Pomeron which is superimposed on a non-leading soft
Pomeron with $ \alpha_{P}(0) $ = 1, plus a Regge contribution. As
 can be seen from the middle section of Table I, we obtain an
excellent fit with $ \frac{ \chi^{2}}{d.f.} $ = 0.96. This is
 demonstrated in Figs. 1-3 where we show the hybrid $ \Omega_{IV} $
 fit compared to the relevant experimental data. The energy scale
was choosen so as to include our predictions for LHC and SCC.
The quality of our fit is even more impressive if we consider the
 fact that some of the ISR data points, at the same energy differ
by more than one standard deviation. The one data point which is not
 in agreement with the fit is the $ \frac{\sigma_{el}}{\sigma_{tot}}$
value as measured by UA4. We call attention to the fact that an
alternative analysis of the UA4 raw data  \cite {HM} yields
a ratio which is compatible with our fit.

\par
A basic conclusion that can be drawn from our analysis, with or
without eikonalization, is that $ \sigma_{tot} $ growth is
associated with both the expansion and the blackening of a(s,b) and
$ \Omega(s,b) $ with s. In the hybrid model we have  \\
$ \frac{\Omega( \sqrt{s} =1800,b=0)}{ \Omega( \sqrt{s} =23,b=0)} $
 = 1.50 and $ \frac{R^{2}( \sqrt{s} =1800)}{R^{2}(\sqrt{s} =23)} $
=1.48.  An identical result was obtained by Chou and Yang \cite {CY}
who found, in the framework of their model, the need to
 change both the normalization and the radius of their $ \Omega(s,b) $
 as a function of s.

\par
The results of the Lipatov 1 sector of the hybrid
 model can be readily  compared with some
theoretical expectations \cite{LR,LERY} in which
\begin{eqnarray}
 \Omega(s,b)= \sigma_{0} \frac{s^{\omega_{0}}}{\sqrt{ln s}}
\exp^{- \frac{b^{2}}{R^{2}(s)}}
\end{eqnarray}
with
\begin{eqnarray}
R^{2}(s) = R^{2}_{0} + 4 \alpha_{eff}^{\prime} \sqrt{lns}
\end{eqnarray}
In the above one should distinguish between parameters that are
determined as a direct consequence of perturbative QCD,
 and those which depend
on the boundary conditions and the LLA used by the authors of references
\cite{LR,LERY}. On the qualitative level our fit
supports the Lipatov energy dependence
$ \frac{s^{a_{2}}}{(lns)^{a_{3}}} $  with both $ a_{2} $ and
$ a_{3} $  positive as expected.  We also obtain
 $ R^{2}_{L1}(s) \propto \sqrt{lns} $, in agreement with references
\cite{LR,LERY}. On the other hand, we see that the actual
numerical values of our fitted $ a_{2},a_{3},a_{4}$ and $a_{5} $
differ from the numbers calculated by references \cite{LR,LERY}.
This difference can be traced to the fact that the above calculations
were made in the LLA with specific boundary conditions. Both may
require some modifications. We find it encouraging that
$ R^{2}_{L1}(\sqrt{s}=23) \approx$ 17 Ge$V^{-2} $, which corresponds
to a radius of $1 fm$. It is interesting to examine both the low and
the high energy limits of the hybrid model. We note that Lipatov 1
in the low energy limit predicts B values which are considerably
smaller than the experimental data. This is compensated, however, by
 the non leading soft sector, producing an overall reproduction of the
 data. This is not suprising, as the hybrid model suggests a smooth,
rather than an abrupt (threshold), continuation from soft to hard
 physics.
Below the black limit,  Lipatov 1 produces a
$ (lns)^{ \frac{3}{2}} $ dependence of $ \sigma_{tot} $ on s.
 This is less than the $ (lns)^{2} $ allowed by the Froissart limit
\cite{FR}. In order to investigate this point further, we have also
 examined an alternative Lipatov 2 parametrization in which
 $ R^{2}_{L1} $ of Eq. (11) is replaced by
\begin{eqnarray}
R^{2}_{L2} = a_{4} +a_{5} \sqrt{lns} + a_{6} lns
\end{eqnarray}
  and we have
\begin{eqnarray}
\Omega_{V}(s,b) = A_{L2}(s,b) + A_{P}(s,b) \pm A_{R}(s,b)
\end{eqnarray}
 This model has an  $ (lns)^{2} $ dependence built in,
provided $ a_{6} \,  >$  0. Our best fit with this form is
 $ \frac{ \chi^{2}}{d.f.} $ = 1.02, is also given in the middle
sector of Table I.
  We find
  $ a_{6} $ = 0.10 wheras $a_{5}$ = 0.789, rather close to the LLA
estimates.

\par
  Both $ \Omega_{IV} $ and $ \Omega_{V} $ models produce a
curvature C which changes sign at $ \sqrt{s} \approx $  1200 GeV
and becomes more negative with increasing s. This result which is
compatible with other models, is just another indication that
 the data examined is far from asymptotia. We note that
the non-eikonalized  models  predict
 very small C values over the entire range
 23 $ \leq \sqrt{s} \leq $ 1800 GeV. Unfortunately, a direct experimental
measurement of C at t = 0 is not possible, and the extrapolation
of C measured at some finite  t  to t=0, is not reliable.
\par As a check of the parametrization obtained in fit $ \Omega_{IV} $
  ( where the parameters were determined from fitting data in
 the energy range
 23 $ \leq \sqrt{s} \leq $ 1800 GeV ), we have also evaluated the
 predictions of the model for
 $ \sigma_{tot}, \frac{\sigma_{el}}{ \sigma_{tot}} $ , and B
 for lower energies i.e.  5 $ \leq \sqrt{s} \leq $ 23 GeV. The results
 are in very good agreement with the experimental data are shown in
 Figs. 1-4.
\par As a further check on our assumption of taking
 Re f(s,t) $ \approx $ 0 for small values of t, we evaluate the
 differential cross-section  obtained from the contribution of Im f(s,t)
only. From the equality Im f(s,0) = $ \frac{ \sigma_{tot}}{ 4 \pi} $
, we take
\begin{eqnarray}
    (\frac{d \sigma}{dt})_{t =0} \approx  \frac{ \sigma^{2}_{tot}}{16
                                                       \pi}
\end{eqnarray}
so that
\begin{eqnarray}
    \frac{ d \sigma}{dt} \approx  \frac{1}{16 \pi} \sigma^{2}_{tot}
 \cdot  \exp (Bt + C t^{2})
\end{eqnarray}
for evaluating the differential cross-section at small t values.
In Fig.5 we show how our  best parametrization describes
the small t - behaviour of the differential elastic cross section. We
 interprete this figure as  support of the approximations used, and
to show how close the physical picture
 we extract from this fitting, is to reality.
\par  We now discuss the real part of the scattering amplitude
which is intimately connected with the crossing symmetry. Our basic
assumption was that the real part is small. We now check
this assumption and calculate the real part for our parametrizations.
To do this, we separate the different contributions of positive
and negative signature in the whole amplitude $ f$ (s, t) ( see Eq. (1)).
In our case, this is easy to do, since the negative signature
contribution is due only to reggeon exchange. Finally
\begin{eqnarray}
   f (s, t ) \,\,=\,\, f_{+}(s, t ) \,\,\pm \,\,f_{-} (s, t )
\end{eqnarray}
for $ pp $ and $ p \bar p $ respectively.
\par
1. The real part of the negative signature contribution
is given by the usual formula, namely
\begin{eqnarray}
   \frac{ Re f _{-} (s, t )}{ Im f_{-} (s, t )}\,\,=
\,\,\tan( \frac{\pi}{2} \alpha_{Reggeon}( t ) )\,\, .
\end{eqnarray}
\par
2. For the positive signature
 we calculated
$Re f_{+} ( s, t )$ as follows
\begin{eqnarray}
   \frac{Re f_{+} ( s, t = 0)}{Im f_{+} (s, t = 0)}\,\,= \,\,- \, cot
( \frac{\pi}{2} \alpha^{eff}_{+} (t = 0 ))\,\,.
\end{eqnarray}
where $ \alpha^{eff}_{+} ( t = 0 ) $  was calculated
 by fitting $f_{+}(s,t = 0) $
 as a power $ s^{\alpha^{eff}_{+} (t = 0 )}$ in the
wide range of energy.
Fig. 6 shows the result of the above calculation for our the best
fit and illustrates that
 $ \rho =  \frac{Re f (s,t = 0)}{ Im f (s,t = 0  )} $ is small
enough to support
 the assumption that one can neglect the real part of the amplitude
for small values of momentum transfer squared.
 One can also see that we get a good
description of the experimental data that have not been included
in the fitting procedure, except of the UA4 data point at $ \sqrt{s} $
= 546 GeV, which has been a suject of much debate \cite{HM}.
 \footnote{
 The knowledge of $Im f(s,t)$ and $\rho$ enables us to calculate
the real part of the amplitude at any $t$ utilizing \cite{MAR}
$$
Re f (s,t ) \,\,=\,\,\rho \frac {d}{dt} ( t Im f (s,t))\,\,.
$$
  The above expression completes our restoration of the crossing
  symmetry in the case when $\rho$ is small enough.}.
\par
   An alternative method for examining the various suggested forms
for the Pomeron is based on the additive quark model \cite{AQM}.
In such an approach we introduce  \cite{LA,LR} two scales
for the hadron: it's size with a radius of $R_{h}$ $ \approx  1 fm$, and
the size of a constituent quark with $ R_{Q} \approx  0.1 - 0.2 fm$.
Based on this slightly naive picture, we can then follow Chou and Yang
\cite{CYY} and write the scattering amplitude
 in the form
\begin{eqnarray}
f(s,t) = 9 G^{2}(t) f_{QQ}(s,t)
\end{eqnarray}
where $ G^{2}(t) = (\frac{\nu^{2}}{\nu^{2} -t})^{2} $ with
 $ \nu^{2}$ = 0.71 $GeV^{2}$, and $f_{QQ}$ denotes the quark-quark
 scattering amplitude. We now apply  our assumptions for the Pomeron
and Regge structures to the QQ amplitude.

\par
We assume a simplified static model \cite{LA,LR} where the
QQ interaction is observed by static spectator quarks. In such a
model, we define $ a^{QQ}(s,b)$ and $ \Omega^{QQ}(s,b)$ in
analogy to Eq.(5), obtaining $ \sigma^{QQ}_{tot}, \sigma^{QQ}_{el}$,
$ B_{QQ}$ and $C_{QQ}$. The hadronic properties are derived from:
\begin{eqnarray}
\sigma_{tot} = 9 \sigma^{QQ}_{tot}  \\
B = B_{QQ} + \frac{8}{\nu^{2}} \\
C = C_{QQ} + \frac{4}{\nu^{4}}
\end{eqnarray}
{}From the relation
\begin{eqnarray}
\sigma_{el} = \frac{ \sigma_{tot}^{2}}{16 \pi} \int dt
e^{Bt +Ct^{2}}
\end{eqnarray}
 we can calculate the ratio $ \frac{ \sigma_{el}}{\sigma_{tot}} $.
We have examined this model in detail, utilizing the various
parametrizations shown before, and were unable to fit the data.
We consistently obtained  $ \frac{\chi^{2}}{d.f.}  > \, $  10,
which is indicative of our inability to reconcile,
 $ \sigma_{tot} $ , the slope B, and the curvature parameter C in such
a model. It has been suggested \cite{LR,GLR}, that the QQ
amplitudes approach the black disc limit. To test this, we have also
examined a parametrization in which
$$a^{QQ}(s,b)\, = \, (1 - \frac{a_{1}}{s^{a_{2}}}) \cdot
\left[\theta[a_{3}lns + a_{4} - b]\right]\,\, + $$
\begin{eqnarray}
+\,\,  \theta[b- a_{3}lns - a_{4}] \cdot
     exp[- \frac{(b-a_{3}lns - a_{4})^{2}}{a_{5}}]
      \pm a_{R}^{QQ}(s,b)
\end{eqnarray}
  Here again, we obtained a $\frac{\chi^{2}}{d.f.} > \, $ 10.
 Our consistent inability to obtain a reasonable fit , is a strong
indication that the additive quark model, at least in the naive form
we have studied, is not amenable to our problem.\par

\par
Finally, we have also examined the possibility that the increase
of $ \sigma_{tot} $ with increasing s is connected to the onset
of minijets produced by parton-parton collisons at energies above the
ISR region \cite {DP}. In such a model the b-space transform of
$ G^{2}(t) $  is normalized by the hard parton-parton cross sections
to which we also add a  soft background.
We disregard the fact that the various authors of reference \cite{DP}
 actually calculated an inclusive rather than a total cross section,
and examine the generic form
\begin{eqnarray}
 \Omega_{VI}(s,b) = [ a_{1} \frac{s^{a_{2}}}{(lns)^{a_{3}}} + a_{4}
                     +  \frac{a_{5}}{\sqrt{s}}] .
            \frac{\nu^{5}b^{3}}{96 \pi}K_{3}(\nu b)
\end{eqnarray}
It is obvious that in this type of model, the increase in
$ \sigma_{tot} $ with s is due to an increase of $ \Omega $ through
blackening rather than expansion, which will only start when
 a(s,b = 0) = 1. Our results for this parametrization are summarized
 at the bottom of Table I. The $ \frac{ \chi^{2}}{d.f.} $ = 3.7
 we have obtained implies that such a model cannot accurately account
for the observed data. We therefore disagree with the conclusions
of Block, Halzen and Margolis \cite {BHM} who have normalized
such a model to the Tevatron data in order to make predictions
for energies in the LHC and SSC domain. Our analysis, which shows
 both a blackening and expansion of a(s,b) and  $ \Omega(s,b) $,
 excludes the possibility that the onset of minijets is the exclusive
 reason for rise of $ \sigma_{tot} $ with energy.

\par
We summarize with the following conclusions and remarks; \\
 1) Our parametrization denoted by $ \Omega_{IV} $ provides a very
 good description of the experimental data over the energy range
5 $ \leq \sqrt{s} \leq $ 1800 GeV. \\
 2) Regardless of details, the data in the
  23 $ \leq \sqrt{s} \leq $ 1800 GeV range shows both moderate
 blackening and expansion of a(s,b). \\
3) The data is compatible with a smooth transition from a soft to
a hard Pomeron contribution  which can account for the rise  of
$ \sigma_{tot} $ with s.
 We  find that the onset
 of hard minijet production cannot exclusively account for the
rise of $ \sigma_{tot} $. Our analysis provides strong evidence
that the Lipatov Pomeron is the proper way  to describe the
rise in $\sigma_{tot}$ which is due to hard processes, such as
minijet production. \\
4) The additive quark model, at least in it's naive form, fails
completely to provide a simultaneous fit to the data.
 \footnote{
  After the completion of this work we learnt of
   a more sophisticated approach
to the  quark model   (see refs.\cite {PU}
  \cite{RY}) that takes into account the interaction
 between more than one quark from each hadron.}        \\
5) The various models we have discussed produce somewhat different
values of  $ \sigma_{tot}$ , B and $ \frac{ \sigma_{el}}{ \sigma_{tot}} $
at $ \sqrt{s} $ =40 TeV. These are summarized in Table II.
We note that our predictions for $ \sigma_{tot}$ are higher than
the 121 mb predicted by reference \cite{BHM} and the 125 mb predicted
by reference \cite{HM}.
It is interesting to check our asymptotic predictions at
the Planck scale  s $ \approx 10^{38} $ Ge$V^{2}$. We observe
that, asymptotically, the various models compared in Table II
have reached the black limit where $ \frac{\sigma_{el}}{\sigma_{tot}}$
= 0.5, and have converged to the same limiting cross section
$ \sigma_{tot}$ = 1010 mb. If we compare this value with the Froissart
 bound \cite{FR}
\begin{eqnarray}
\sigma_{tot} = \frac{\pi}{ \mu^{2}} ln^{2} \frac{s}{s_{0}}
\end{eqnarray}
 we find that $ \mu \approx $ 3 GeV for $ \sqrt{s_{0}} <$ 23 GeV.\\

\vspace{5 mm}
\par
{\bf Acknowledgements:}
Part of this work was carried out while E.G. and U.M. were visiting
DESY, they wish to thank the MINERVA Foundation for its support.
 We wish to thank the Theory Group  for their kind hospitality.
Stimulating discussions with J. D. Pumplin are gratefully acknowledged.

\newpage
\textwidth 15.4 cm

\newpage

\vspace{3 mm}
\begin{scriptsize}

\begin{center}
{{\bf Table 1 :} Values of parameter for different
          parameterizations of the Pomeron.}
\end{center}
\end{scriptsize}

\newpage
\section*{Figure captions}
{\bf Figure 1} $\;\;$  $\sigma_{tot}$  in mb vs $ \sqrt{s} $ in GeV.
The measured cross sections are from \cite{DATA} .
 The dashed (solid) line are the predictions for pp ( $ p {\bar p} $ )
 for fit $\Omega_{IV}$.   \\
\\
{\bf Figure 2} $\;\;$ B, the measured and predicted nuclear slope
parameter in units of $( GeV )^{-2} $
 vs $ \sqrt{s} $ in GeV.  Data from \cite{DATA}. Dashed (solid) curve
 as in Fig. 1. \\
\\
{\bf Figure 3} $\;\;$ The ratio $ \frac{\sigma_{el}}{\sigma_{tot}}$
vs. $\sqrt{s} $ in  GeV.
Data from \cite{DATA} . Dashed (solid) curve as in Fig. 1.\\
\\
{\bf Figure 4} $\;\;$ The predicted value of the curvature parameter
 C in units of $( GeV)^{-4}$  vs. $\sqrt{s} $ in  GeV.
Dashed (solid) curve as in Fig. 1.\\
\\
{\bf Figure 5} $\;\;$ Elastic differential  ${\bar p} p $ cross section
at small -t , at representative energies. Data from  \cite {DATA} . \\
\\
{\bf Figure 6} $\;\;$  Ratio of real to imaginary parts  of forward
elastic amplitude
vs. $\sqrt{s} $ in  GeV. Data from \cite {DATA}.
Dashed (solid) curve as in Fig. 1.\\
\\
\end{document}